\documentclass[12pt,a4paper]{article}
\usepackage{amsfonts}
\usepackage{amsmath}
\usepackage{amssymb}
\usepackage{hyperref}

\usepackage{color}

\newtheorem{theorem}{Theorem}

\newtheorem{lemma}[theorem]{Lemma}

\newfont{\cmss}{cmss10}
\newcommand{\ssd}{\mbox{{\cmss C}\hspace{-2mm}{\tt +}}}

\newfont{\cms}{cmss12}
\newcommand{\psd}{\mbox{{\cmss C}\hspace{-1.8mm}\raisebox{0.3mm}{\cms
x}}}

\newcommand{\qed}[0]{\textbf{QED}}

\date{December 2005, Version 2006.12.24}
\author{Erik Taflin\footnote{taflin@eisti.fr;  EISTI,
Ecole International des Sciences du Traitement de l'Information,
Avenue du Parc, 95011 Cergy, France} \footnote{Acknowledgement: The author thanks
Avy Soffer for pointing out the interest of the particular cases (\ref{Eq1.3})
and (\ref{Eq1.4})  of the NLKG (\ref{Eq1.2}).}}
\title{Simple Non Linear Klein-Gordon Equations in $2$ space dimensions, with long range
scattering}

\begin{document}
\maketitle

\begin{abstract}
We establish that solutions, to the most simple NLKG equations in $2$ space dimensions
with mass resonance,
exhibits long range scattering phenomena. Modified wave operators and solutions
are constructed for these equations. We also show that the modified wave operators
can be chosen such that they linearize the non-linear representation of the Poincar\'e
group defined by the NLKG.
\end{abstract}
\section{Introduction}
The purpose of this article is to study  Non-Linear Klein-Gordon Equations
in $2$ space dimensions with a finite number of masses $m_i>0,$ having a mass
resonance of the following kind,
intrduced in \cite{S-T 85}:
For some $j, \ j_1, \ j_2,$ there exists numbers
$\epsilon_{j_1}, \ \epsilon_{j_2}=\pm 1$
such that
\begin{equation}   \label{Eq1.1}
m_j= \epsilon_{j_{1}} m_{j_1} + \epsilon_{j_2} m_{j_2}  \ .
\end{equation}
The equations for the real valued functions $\varphi_{i}$ are:
\begin{equation}
 \label{Eq1.2}
(\square + m_i^{2}) \varphi_{i}= F_i (\varphi, \partial \varphi),
\end{equation}
where
$\varphi$ is the vector with components $\varphi_{i},$
$t \in  \mathbb{R},$ $x \in \mathbb{R}^{2},$ $ \varphi_{i}(t,x) \in \mathbb{R},$
$\partial = (\partial_{0},\partial_{1},\partial_{2}),$
$\partial_{0} =\frac{\partial}{\partial t},$
$\partial_{j} = \frac{\partial}{\partial x_{j}}$ for $j=1,2,$
$\Delta = {\sum_{i=1}^{2} \partial_{i}^{2}},$
$\square = (\partial_{0})^{2} - \Delta .$
The $F_i$ are real $C^{\infty}$ functions, vanishing together with their first derivative
at the origin.

In this paper we shall study the simplest cases of eq. (\ref{Eq1.2}),
when condition (\ref{Eq1.1}) is satisfied.
For a given mass $ m>0 ,$ we consider the following  two systems of non-linear
Klein-Gordon (NLKG) equations, each containing one of the basic critical terms of
(\ref{Eq1.2}): %
\begin{equation}
 \label{Eq1.3}
(\square + m^{2}) \varphi_{1}= 0, \quad
(\square + (2m)^{2}) \varphi_{2}=(\varphi_{1})^{2}
\end{equation}
and
\begin{equation}
\label{Eq1.4}
(\square + m^{2}) \varphi_{1}=\varphi_{1} \varphi_{2},  \quad
(\square + (2m)^{2}) \varphi_{2}=0.
\end{equation}
It easily follows that the Cauchy problem for each of the system of equations
(\ref{Eq1.3}) and (\ref{Eq1.4}) has global
solutions for large initial data (see Theorem \ref{th1} for a precise formulation).
The scattering problem is more interesting, since it is only the quadratic terms in (\ref{Eq1.2})
which can give rise to long-range phenomenas:

1) We establish (Theorem \ref{th2}) that the systems
(\ref{Eq1.3}) and (\ref{Eq1.4}) have ``long range'' modified wave operators and that
they fail to have ``short range'' wave operators.
This is due to the second degree ``mass resonance'', defined by (\ref{Eq1.1}),
which is present in these systems together with the $t^{-1}$ time decrease of
the $L^{\infty}(\mathbb{R}^{2})$-norm of solutions of the linear K-G equation.
This should be compared with the small-data Cauchy and scattering problem for the NLKG
\begin{equation}  \label{Eq1.5}
(\square + m^{2}) \varphi= %
                F(\varphi, \partial \varphi),
\end{equation}
with only one mass $m>0.$ For $n \geq 2$ space dimensions, the scattering theory
of (\ref{Eq1.5}) is short range \cite{S-T 92}
(see also \cite{Horm97} and references therein for further developments),
which reflects the fact that there is no second degree ``mass resonance''.
However, there is a third degree ``mass resonance'' which for $n=1$ together with
the $t^{-1}$ time decrease of the $L^{\infty}(\mathbb{R})$-norm of $\varphi^{2}$
gives rise to the ``long range''
behavior treated in \cite{D01}, \cite{L-S1} and \cite{L-S2} for the cubic NLKG.
We note that the asymptotic completeness of the modified wave operators for
(\ref{Eq1.4}) is not studied in this paper.
The methods in
 \cite{L-S2}, adapted to spaces of initial conditions like
Schwartz spaces, seem to give a promising departure for such future studies.
For (\ref{Eq1.3}) the asymptotic completeness is a trivial
consequence of Theorem \ref{th1} and Theorem \ref{th2}.

2) For $n \geq 1$ space dimensions, all formal nonlinear representations of the
Poincar\'e group only involving massive fields are (at least formally) linearizable
(see \cite{T 84} where the corresponding cohomology was proved to be trivial).
Then a natural question is:
can modified wave operators be chosen such that they
intertwine the non-linear representation, of the Poincar\'e group %
(and its Lie algebra) naturally defined on initial conditions for (\ref{Eq1.3}) and (\ref{Eq1.4}),
and the linear representation defined by their linear part %
i.e.
\begin{equation}
\label{Eq1.5.1}
(\square + m^{2}) \varphi_{1}=0,  \quad
(\square + (2m)^{2}) \varphi_{2}=0.
\end{equation}
We prove that the answer is yes (Theorem \ref{th2}). This is not at all automatic.
For example, %
it is not possible for the Maxwell-Dirac equations in three space dimensions.
 In fact, as was proved in
\cite{FST97}, MD is non-linearizable, on natural spaces of initial conditions.

We next write equations (\ref{Eq1.3}) and (\ref{Eq1.4}) as evolution
equations in a Hilbert space $E.$ The variable
$a(t)=(a_{1,+}(t), a_{1,-}(t),a_{2,+}(t), a_{2,-}(t))$ is defined by:
\begin{equation}\label{a1.4}
a_{j,\epsilon} (t) = \dot{\varphi}_{j} (t)+ \epsilon i \omega_{jm}
(-i \nabla) \varphi_{j}(t),\;\; \epsilon = \pm 1,
\end{equation}
where $\omega_M (p)=(M^2+|p|^2)$ and
$\dot{\varphi}_{j}(t,x) = \frac{\partial}{\partial t}\varphi_{j} (t,x)$.
The inverse of the transformation (\ref{a1.4}) is
\begin{equation}\label{a1.5}
\varphi_{j}(t) = (2i\omega_{jm}(-i \nabla))^{-1} (a_{j,+}(t)-a_{j,-}(t)),
\quad \dot{\varphi}_{j}(t) = 2^{-1}(a_{j,+}(t)+a_{j,-}(t)).
\end{equation}
Equations (\ref{Eq1.3}) and (\ref{Eq1.4}) then reads
\begin{equation}
\begin{cases}
\label{a1.6}
&\frac{d}{dt} a_{1}(t) = i\omega_{m}(-i \nabla) (a_{1,+}(t),-a_{1,-}(t))
+ (F_{1}(a(t)),F_{1}(a(t))) \\
&\frac{d}{dt} a_{2}(t) = i\omega_{2m}(-i \nabla) (a_{2,+}(t),-a_{2,-}(t))
+ (F_{2}(a(t)),F_{2}(a(t))),
\end{cases}
\end{equation}
where in the case of equation (\ref{Eq1.3})
\begin{equation}\label{a1.7.1}
\begin{split}
F_{1}=0, \; \;
F_{2}(a(t)) =&\left((2i\omega_{m}(-i\nabla))^{-1}(a_{1,+}(t)-a_{1,-}(t))\right)  \\ &
        \quad \left((2i\omega_{m}(-i\nabla))^{-1}(a_{1,+}(t)-a_{1,-}(t))\right)
\end{split}
\end{equation}
and in the case of equation (\ref{Eq1.4})
\begin{equation}\label{a1.7.2}
\begin{split}
F_{2}=0, \;\; 
F_{1}(a(t)) =&\left((2i\omega_{m}(-i\nabla))^{-1}(a_{1,+}(t)-a_{1,-}(t))\right)  \\ &
      \quad   \left((2i\omega_{2m}(-i\nabla))^{-1}(a_{2,+}(t)-a_{2,-}(t))\right).
\end{split}
\end{equation}
The real Hilbert space $E$ is defined by $E=E_{(1)} \oplus E_{(2)}$ with norm
$\|f\|_E =(\sum_{j=1,2} \|f_j\|_{E_{(j)}})^{1/2},$ where
$E_{(j)}$ is the real subspace of $E_{(j)}^{C}=E_{(j,+)} \oplus E_{(j,-)}$ such that the image
of the transformation (\ref{a1.5}) only contains  real functions. The norms in
the complex Hilbert spaces $E_{(j)}^{C}$ and  $E_{(j,-)}$ are given by
\begin{equation} \label{Eq1.5.1.-1}
\|f_j\|_{E_{(j)}}=(\sum_{\epsilon = \pm} \|f_{j,\epsilon}\|_{E_{(j,\epsilon)}})^{1/2}
\; \mathrm{and} \;
\|f_j\|_{E_{(j,\epsilon)}} =%
\|(\omega_{jm} (-i\nabla))^{-1/2}f_{j,\epsilon}\|_{L^{2}}.   %
\end{equation}
We shall define modified out and in wave operators
$\Omega_{+}: \mathcal{O}^{+} \rightarrow \mathcal{O}^{0}$
and $\Omega_{-}: \mathcal{O}^{-} \rightarrow \mathcal{O}^{0}$ respectively,
by introducing, for given scattering data  $f \in \mathcal{O}^{\delta},$ %
$\delta =\pm,$  an approximate solution $a^{(\delta)}(f)$ satisfying for some
initial condition $a(0)$ of equation (\ref{a1.6}) and for $\alpha =0:$
\begin{equation}
\label{Eq1.5.3}
\lim_{t \rightarrow \delta \infty} (1+|t|)^\alpha \| a(t)-(a^{(\delta)}(f))(t) \|_{E} =0.
\end{equation}
By the uniqueness of the solution $a$ we can now define
\begin{equation}
\label{Eq1.5.4}
 a(0)=\Omega_{\delta}(f).
\end{equation}
Since the cases $\delta = \pm$ are so similar,
we limit ourselves to $\delta =+.$ A study of the large time behavior of solutions
of (\ref{a1.6}) by stationary phase methods and the use of \cite{T 84} to
construct linearization maps of nonlinear representations of the Poincar\'e group
leads to a choice of approximate solutions $a^{(+)}(f).$
With the notation
$V(t)_{(j,\epsilon)}=\exp(   i\omega_{jm}(-i \nabla) t  )$ %
we define $ (a^{(+)}(f))(t) = V(t)(b^{(+)}(f))(t),$ where
in the case of (\ref{a1.7.1})\footnote{The Fourier transformation $f \mapsto \hat{f}$ is here defined by
$\hat{f} (k) = (2 \pi)^{- 1}   \int_{\mathbb{R}^2}   e^{-ikx}
f(x)   dx.$}
\begin{equation}
\begin{cases}
\label{a1.6+}
& b^{(+)}_{1}(t) =  f_{1} \\
&(b^{(+)}_{2,\epsilon}(t))^{\hat{}}(k) =
\hat{f}_{2,\epsilon}(k)
   - i\epsilon \ln{(1+\frac{t(2m)^{2}}{\omega_{2m}(k)}}) \, \frac{1}{8m} (\hat{f}_{1,\epsilon}(k/2))^{2}
\end{cases}
\end{equation}
($f$ in $(b^{(+)}(f))(t)$ has here been omitted) and in the case of (\ref{a1.7.2})
\begin{equation}
\begin{cases}
\label{a1.7+}
& b^{(+)}_{1}(t) = \exp{\left(\frac{1}{4m} L(f_{2})
    \ln{(1+t m^{2} (\omega_{m}(-i \nabla))^{-1}}) \right)}f_{1} \\
&
b^{(+)}_{2}(t) = f_{2},
\end{cases}
\end{equation}
where for $g \in E_{(1),\infty}^{C}$ and  $h \in E_{(2),\infty}^{C},$
 $E_{(j),\infty}^{C}
=S(\mathbb{R}^{2},\mathbb{C}) \oplus S(\mathbb{R}^{2},\mathbb{C}),$
\begin{equation} \label{Eq1.8}
((L(h)g)_\epsilon)^{\hat{}}(k) \equiv (L_\epsilon(h)g_{-\epsilon})^{\hat{}}(k)
=i \epsilon \hat{h}_\epsilon (2k) \hat{g}_{-\epsilon}(-k).
\end{equation}
Then, for a given $f \in E_\infty =E_{(1),\infty} \oplus E_{(2),\infty},$
where $E_{(j),\infty}=E_{(1)} \cap E_{(1),\infty}^{C},$
$a$ is formally a solution of
\begin{equation} \label{Eq1.9}
a(t)= (a^{(+)}(f))(t) -\int_t^\infty V(t-s) \left(T^{2}_{P_0} (a(s))
           -V(s) (\dot{b}^{(+)}(f))(s) \right) \, ds,
\end{equation}
where $\dot{b}^{(+)}(f))(t)=\frac{d}{dt}(b^{(+)}(f))(t)$ and see (\ref{a1.11}) for $T^{2}_{P_0}.$
A rigorous study of this equation in next section, will lead to the construction and covariance
properties of modified wave operators (Theorem \ref{th2}).

The construction of modified wave operators and solutions of more general evolution equations
also leads to an equation analog to (\ref{Eq1.9}), where the recipe
(usually based on an iteration starting with a free solution)  how to find
an approximate solution $a^{(+)}(f)$ for given scattering data $f$ has to be specified
in each particular case. In the case of relativistic covariant equations,
this was accomplished for the MD eq.
in three space dimensions  \cite{FST87} (see also \cite{FST97}) for asymptotic completeness)
and for NLKG in one space dimension \cite{D01}, \cite{L-S1} and \cite{L-S2}.
For NLS it was accomplished in  \cite{0-91}  and for several other non-relativistic equations
in \cite{GB03} and references therein to related papers by the same authors.

The Poincar\'e group  $\mathcal{P}=\mathbb{R}^{3}\psd SO(2, 1)$
acts on elements $y=(y^{0},y^{1},y^{2})$ in the 3-dimensional Minkowski space by
$g y = \Lambda y - a,$ where $g = (a,\Lambda),$ $\Lambda \in SO(2, 1)$ and $a \in \mathbb R^2.$
$\mathcal{P}$ acts on real functions $f$ on the Minkowski space by a linear representation
$R:$ %
\begin{equation}\label{a1.7.3}
(R_g f)( y ) = f (g^{-1} y), \quad y \in \mathbb R^3.
\end{equation}
Covariance of the NLKG under the representation $R$ leads to nonlinear
representations of $\mathcal{P}.$
$\Pi = \{P_0,P_1,P_2,R,N_{1},N_{2}\}$ denotes an ordered standard basis of the Poincar\'e Lie algebra
$\mathfrak{p} = \mathbb{R}^{3}\ssd so(2,1)$ in $3$ dimensions.
Here $P_0,$ $P_1,$ $P_2,$ $R,$ $N_{1}$ and $N_{2}$ are respectively, the time translation,
the two space translation, the space rotation and the two boost generators.
We define a linear representation $T^{1}$ of
$\mathfrak{p}$ in the Schwartz space $E_{\infty}$
of elements $f=(f_{1,+},f_{1,-},f_{2,+},f_{2,-})$ by:
\begin{align}\label{a1.8}
&(T^{1}_{{P}_{0}}f)_j  =i\omega_{jm}(-i\nabla) (f_{j,+} , - f_{j,-}) , \quad j=1,2, \\
&T^{1}_{{P}_{n}} f  = {\partial}_{n} f ,\quad  n=1,2 ,  \\
&T^{1}_{R}f = m_{12}f ,\quad
m_{12} = x_{1} {\partial}_{2} -x_{2} {\partial}_{1} , \\
&(T^{1}_{N_{n}}f)_j (x)  = (i\omega_{jm}(-i \nabla)
x_{n} f_{j,+} , -i\omega_{jm}(-i \nabla) x_{n} f_{j,-}) , \;  j,n=1,2.
\end{align}
The non-linear representation $T$  of $\mathfrak{p}$ on
$E_{\infty}$ (see \cite{FSP77}), is obtained by the fact that equations
(\ref{Eq1.3}) and (\ref{Eq1.4})
are manifestly covariant:
\begin{equation}\label{a1.10}
T_{X} = T^{1}_{X} + T^{2}_{X} ,\quad  X \in \mathfrak{p} ,
\end{equation}
where %
for $f \in E_{\infty}$ the quadratic term $T^{2}$ is given by
\begin{align}\label{a1.11}
&T^{2}_{P_0} (f) =  (F_1(f) , F_1(f),F_2(f) , F_2(f)), \\
&T^{2}_{P_1} =T^{2}_{P_2}=T^{2}_{R} = 0, \\
&(T^{2}_{N_{n}} (f))(x) = x_{n}(T^{2}_{P_0} (f))(x) ,\quad  n=1,2.
\end{align}
In particular, equation (\ref{a1.6}) reads
\begin{equation}
\label{a1.6.1}
\frac{d}{dt} a(t) = T_{P_0} (a(t).
\end{equation}
The representation $T^{1}$ is the differential of a unitary representation $U^{1}$ of
the Poincar\'e group $\mathcal{P}$ in the Hilbert space $E.$
Let $\Pi'$ be the standard basis of the universal enveloping algebra $\mathcal{U}(\mathfrak{p})$
of $\mathfrak{p}$ corresponding to $\Pi.$  We give $\Pi'$ its lexicographic
order with respect to the ordered basis $\Pi.$
Let $\vert Y \vert$ be the degree of $Y \in \Pi'.$ %
The space $E_n$ of n-differentiable vectors for the representation
$U^{1}$ in $E$ coincides with the Hilbert space obtained by the completion of $E_\infty$ 
with respect to the norm (summing over $Y \in \Pi'$ and $\vert Y\vert \leq n$)
\begin{equation}
\Vert  u\Vert _{E_n} = %
 (\sum \Vert  T^1_Y u \Vert ^2_E)^{1/2}, 
\end{equation}
where $T^1_Y$ %
is defined by the canonical extension of $T^1$ from $\mathfrak{p}$ to
$\mathcal{U}(\mathfrak{p}).$  %
We have $E_\infty \subset E_j \subset E_i \subset E_0 = E$ for $i\leq j$.
$U^1_j$ and $T^1_j$ denote the representations obtained by restricting
$U^1$ and $T^1$ to $E_{(j)}$ and $E_{(j), \infty}$ respectively.
Here $E_{(j), n}$ and $E_{(j), \infty}$ denotes the image of the canonical
projection of $E_{n}$ and $E_{ \infty}$ on $E_{(j)}.$ We note the well-known fact
that the norms $\Vert  \;\; \Vert _{E_n}$ and $q_n$ are equivalent,
where (summing over multi-indices with $x^\mu=x_1^{\mu_1}x_2^{\mu_2}$
and $\nabla^\mu=\partial_1^{\mu_1}\partial_2^{\mu_2}$)
\begin{equation} \label{Eq1.5.1.0}
q_n (f)=\bar{q}_n ((I-\Delta)^{-1/4}f) \; \; \text{and} \;\;
\bar{q}_n (f) = (\sum_{\vert \mu \vert, \, \vert \nu \vert \leq n}
\Vert  x^\mu  \nabla^\nu f \Vert^2_{L^2})^{{1 / 2}}.
\end{equation}

The linear map $ X \mapsto T_X,$
from $\mathfrak{p}$ to the vector space of all $C^{\infty}$ maps from $E_\infty$
to $E_\infty,$  extends to $\mathcal{U}(\mathfrak{p})$ by defining inductively
(see \cite{S-T 92}):
$T_{\mathbb{I}} = I$,
where $\mathbb{I}$ is the identity element in the enveloping algebra, and
\begin{equation}
\label{Eq1.5.1.1}
T_{YX} = D T_Y.T_X, \quad X \in \mathfrak{p},
\end{equation}
where $(DA.B)(f)=(DA)(f;B(f))$ is the the Fr\'echet derivative of the map $A$ at the point $f$ in the
direction $B(f).$ Suppose for the moment that the nonlinear Lie algebra representation
$ X \mapsto T_X$ is (locally) integrable, i.e. in this case
$\forall X \in \mathfrak{p}$ and $\forall f \in E_\infty$ there exists $c>0$
such tat for $|t| < c$
\begin{equation}
\label{Eq1.5.1.2}
\frac{d}{dt} U_{g(t)}(f)=T_{X}(U_{g(t)}(f)), \quad g(t)=\exp{(tX)}.
\end{equation}
Then, for an element $Y \in \mathcal{U}(\mathfrak{p})$  (see \cite{S-T 92}, \cite{S-T 95})
\begin{equation}
\label{Eq1.5.2}
\frac{d}{dt} T_{Ad_{g(t)}(Y)} (U_{g(t)} (f)) = T_{X Ad_{g(t)}(Y)} (U_{g(t)} (f)),
\end{equation}
where the adjoint representation is given by
\begin{equation}
\frac{d}{dt} Ad_{g(t)} Y = [X, Ad_{g(t)} Y] , \quad Ad_{g(0)}Y=Y.
\end{equation}
\section{Main Results}
Since the equation given by (\ref{a1.6}) for $a_2$ (resp. $a_1$) in the case of
(\ref{a1.7.1}) (resp. (\ref{a1.7.2})) is simply a linear K-G, with an inhomogeneous
(resp. linear potential) term, we easily prove the following theorem: %
\begin{theorem} \label{th1} i) There exists $N_0$ such that
for $N \geq N_0,$ $T$ is integrable to a unique global nonlinear analytic group representation
$U$ of $\mathcal{P}$ on  $E_N$ and
$U: \mathcal{P} \times E_\infty \rightarrow E_\infty$   is $C^\infty.$ \\
ii) For all initial conditions  $f \in E_\infty,$ equation (\ref{a1.6}) has a
unique $C^\infty$ solution $a: \mathbb{R} \rightarrow E_\infty.$ \\
iii) For all initial conditions
$(\varphi_{1}(0),\dot{\varphi}_{1}(0),\varphi_{2}(0),\dot{\varphi}_{2}(0))
      \in S(\mathbb{R}^{2},\mathbb{R}^{4}),$ there exists a unique solution
$(\varphi_{1},\varphi_{2}) \in C^{\infty}(\mathbb{R}^{3}, \mathbb{R}^{2})$ of eq. (\ref{Eq1.3})
(resp. (\ref{Eq1.4})). %
\end{theorem}
\textbf{Outline of proof:} Proceeding as in \cite{S-T 92} and \cite{S-T 95},
for $Y \in \mathcal{U}(\mathfrak{p})$ and $X \in \mathfrak{p}$ introduce
$$u_Y (t)= T_{Ad_{\exp{(t X)}}(Y)} (u(t)).$$
Let $u(0)=f \in E_\infty.$ According to equation (\ref{Eq1.5.2})
\begin{equation}
\label{Eq2.1}
\frac{d}{dt}u_Y (t) = u_{X Y}(t), \quad u_Y (0)=T_{Y} (f).
\end{equation}
Let $\mathbb{I} < Y_1 < \ldots < Y_{c(k)}$ be 
the lexicographic ordering of the set of $Y\in \Pi'$ such that $\vert Y\vert \leq k,$ and let 
\begin{equation} \label{Eq2.2}
v_N (t) = (u_{Y_0} (t), u_{Y_i} (t) , \ldots, u_{Y_{c(N)}} (t)), N \geq 0.
\end{equation}
According to formula (2.23a) of \cite{S-T 95}, (\ref{Eq2.1}) leads to an equation for $v_N,$
\begin{equation} \label{Eq2.3}
v_N (t) = U^1_{\exp{(tX)}} v_N (0) + \int^t_0 U^1_{\exp{((t-s)X)}} G_N (v_N (s)) ds,
\end{equation}
for some, in this case, quadratic forms $G_N$ depending on $X.$
We define, for a function $d: \Pi' \rightarrow E$ and for $n \in \mathbb{N}:$
\begin{equation} \label{Eq2.4}
\mathcal{P}_n (d) = (\sum_{\substack{Y\in \Pi' \\ \vert Y\vert \leq n}} \Vert  d_Y \Vert_E^2)^{1/2}.
\end{equation}
Choosing $N_0$ sufficiently large, one obtains from (\ref{a1.7.1}), (\ref{a1.7.2})
and  (\ref{Eq2.3}) using the unitarity of $U^1$ that
\begin{equation} \notag
\mathcal{P}_N (u(t)) \leq \mathcal{P}_N (T(f))
+ \int_{0}^{t}C_{N} \mathcal{P}_N (T(f)) (1+s)^{-1} \mathcal{P}_N (u(s))\,ds, \; N \geq N_0.
\end{equation}
Then by Gr\"onwall's lemma
$\mathcal{P}_N (u(t)) \leq \mathcal{P}_N (T(f)) (1+t)^{C_{N} \mathcal{P}_N (T(f))} < \infty$
 for $t \geq 0.$
Statement (i) now follows by using Theorem 6 of  \cite{S-T 95}.
Statements (ii) and (iii) are direct consequences of (i). \qed

The following two lemmas %
give time decrease of %
 $b^{(+)}(t)$ and its derivatives.
\begin{lemma} \label{lm1}
Let $f \in E_\infty.$ %
Then $t \mapsto b^{(+)}(t)$ is a $C^\infty$ mapping from $[0, \infty [ \; $
to $E_\infty$ and there exists constants $C$ independent of $f$ and $C_{N,n}$ such that for all
$f \in E_\infty,$ $t \geq 0,$ $n \geq 0$ and $N \geq 2$ \\
i) if $F$ is given by (\ref{a1.7.1})  then, $C_{N,n}$ is independent of $f,$
\begin{equation} \label{Eq2.5}
\Vert b^{(+)}_2 (t) \Vert_{E_{(2), N}} \leq \Vert f_2 \Vert_{E_{(2), N}} + C_{N,0} \ln{(2+tm)}
       \|f_1 \|_{E_{(1), N}} \|f_1 \|_{E_{(1),2}}
\end{equation}
and for $n \geq 1$
\begin{equation} \label{Eq2.6}
\Vert \frac{d^n}{dt^n}b^{(+)}_2 (t) \Vert_{E_{(2), N}} \leq C_{N,n} (1+t)^{-n}
       \|f_1 \|_{E_{(1), N}} \|f_1 \|_{E_{(1),2}}
\end{equation}
ii) if $F$ is given by (\ref{a1.7.2}) then, %
$C_{N,n}$ only depends on $\|f_2 \|_{E_{(2), 3}}$ and
\begin{equation} \label{Eq2.7}
\begin{split}
\Vert \frac{d^n}{dt^n}b^{(+)}_1 (t)& \Vert_{{E_{(1), N}}} \leq C_{N,n} (1+t)^{C \|f_2 \|_{E_{(2),1}} -n}
(\ln{(2+tm)})^{2N+1} \\
& (\|f_1 \|_{E_{(1), N}}+ \|f_1 \|_{E_{(1), 3}} \|f_2 \|_{E_{(2), N}}
+\|f_1 \|_{E_{(1), N}} \|f_2 \|_{E_{(2), 3}}).
\end{split}
\end{equation}
\end{lemma}
\textbf{Proof:} We only consider the more difficult case (ii). Expression (\ref{a1.7+})
gives
\begin{equation} \label{Eq1 prooof lm1}
(b^{(+)}_{1}(t))_{\epsilon}^{\hat{}}(k) =
\cosh{(S(t,k))}f_{1,\epsilon}^{\hat{}}(k)
+\frac{\sinh{(S(t,k))}}{S(t,k)}T(t,k),
\end{equation}
where, for given $\epsilon,$
$S(t,k)=(1/4m) |f_{2,\epsilon}^{\hat{}}(2k)| \ln{(1+\frac{t m^{2}}{\omega_{m}(k)})}$
and $T(t,k)=\frac{i \epsilon}{4m}f_{2,\epsilon}^{\hat{}}(2k)
  \ln{(1+\frac{t m^{2}}{\omega_{m}(k)})} f_{1,-\epsilon}^{\hat{}}(-k).$
Let $F_r(z)=\sum_{n \geq 0} z^n / ((2n+r)!),$ $r=0,1.$ %
Then $F_0(z^2)=\cosh{(z)}$ and  $F_1(z^2)=\sinh{(z)}/z.$ The $n$-th derivative satisfies
$|F_r^{(n)}(z)| \leq F_r^{(n)}(|z|)$ and $F_r^{(n+1)}(x) < F_r^{(n)}(x),$ $x\geq 0.$
We define the norms $Q_N$ and $Q'_N,$ $N \geq 1$ by
\begin{equation} \label{Eq2 proof lm1}
Q_N (a) =\|a\|_{L^\infty}+Q'_N (a),
\;\;
Q'_N (a) = (\sum_{\substack{0 \leq \vert \mu \vert \leq N \\ 1 \leq \vert \nu \vert \leq N}}
\Vert \frac{x^\mu}{(1+|x|^2)^{1/4}}  \nabla^\nu a \Vert_{L^2}^2)^{{1 / 2}}.
\end{equation}
Let $h_r(t,k)=F_r(g(t,k)),$ where $g(t,k)=(S(t,k))^2.$
Applying $k^\alpha (\partial/\partial k)^\beta$ on $h_r(t,k),$ for multi-indices
$\alpha$ and $\beta,$ and using the above properties of $F_r,$ the expression
(\ref{Eq1.5.1.0}), Plancherel's theorem and interpolation, give for $N \geq 3:$
\begin{equation} \notag %
Q_N ( h_r(t,\cdot)) \leq C_N F_r(\| g(t,\cdot)\|_{L^\infty})  (1+Q'_{3}(g(t,\cdot)))^{N-1}
(1+Q'_{N}(g(t,\cdot)). %
\end{equation}
We note that
$\| g(t,\cdot)\|_{L^\infty} \leq C^2 \|f_2\|_{E_{(2),1}}^2  (\ln{(2+t m)})^2,$
and that by interpolation
$Q'_{N}(g(t,\cdot)) \leq C'_N  Q'_{N}((\hat{f}_{2})^2) (\ln{(2+t m)})^2
\leq C''_N \|f_2\|_{E_{(2),3}} \|f_2\|_{E_{(2),N}} (\ln{(2+t m)})^2.$ %
Since $F_r(x^2) \leq e^x,$ $x \geq 0,$ we obtain for $ N \geq 3:$
\begin{equation} \label{Eq4 proof lm1}
Q_N (h_r(t,\cdot)) \leq C_N (1+tm)^{C \|f_2\|_{E_{(2),1}}} (\ln{(2+t m)})^{2 N}
(1+\|f_2\|_{E_{(2),3}})^{N-1}(1+\|f_2\|_{E_{(2),N}}).
\end{equation}
Interpolation then gives, with $(H_0(t))^{\hat{}}(k)=h_0(t,k) f_{1,\epsilon}^{\hat{}}(k)$
and $(H_1(t))^{\hat{}}(k)=h_1(t,k) T(t,k):$
\begin{equation} \label{Eq5 proof lm1}
\begin{split}
&\|H_0(t)\|_{E_{(1),N}}+\|H_1(t)\|_{E_{(1),N}}
 \leq   C_N (1+tm)^{C \|f_2\|_{E_{(2),1}}} (\ln{(2+t m)})^{2 N +1} \\ &(1+\|f_2\|_{E_{(2),3}})^{N}
(\|f_1\|_{E_{(1),N}}(1+\|f_2\|_{E_{(2),3}})+\|f_1\|_{E_{(1),3}}\|f_2\|_{E_{(2),N}}),
\end{split}
\end{equation}
which proves (\ref{Eq2.7}) in the case of $n=0.$ Repeated use of
\begin{equation} \label{Eq6 proof lm1}
\frac{d}{dt}b^{(+)}_{1}(t) =\frac{1}{4} L(f_{2})
    (\omega_{m}(-i \nabla)/m+t m)^{-1}b^{(+)}_{1}(t)
\end{equation}
and interpolation leads to, for $N\geq 3$ and $n \geq 1:$
\begin{equation} \label{Eq7 proof lm1}
\begin{split}
&\Vert \frac{d^n}{dt^n}b^{(+)}_1 (t) \Vert_{{E_{(1), N}}} \leq   C_{N,n} (1+tm)^{-n} \\
&(1+\|f_2\|_{E_{(2),3}})^{n-1}
(\|b^{(+)}_1 (t)\|_{E_{(1),N}}\|f_2\|_{E_{(2),3}} +\|b^{(+)}_1 (t)\|_{E_{(1),3}}\|f_2\|_{E_{(2),N}}).
\end{split}
\end{equation}
The case $n=0$ of (\ref{Eq2.7}) and inequality (\ref{Eq7 proof lm1}) prove statement (ii)
of the lemma. \qed
\begin{lemma} \label{lm2}
For all $f \in E_\infty,$ $t \geq 0$ and $n,N \geq 0$ %
there exists a constant $C$ independent of $f,$ and constants $C_{N,n}$ and $N'$ such that \\
i) if $F$ is given by (\ref{a1.7.1}) then, $C_{N,n}$ is independent of $f$ and
\begin{equation} \label{Eq2.8}
\begin{split}
q_N&\left(\frac{d^n}{dt^n}
\left(e^{-i \epsilon \omega_{2m}(-i\nabla)t}
       ((2i\omega_{m}(-i\nabla))^{-1}a^{(+)}_{1,\epsilon}(t) )^{2}
           - \dot{b}^{(+)}_{2,\epsilon} (t)\right)\right) \\
& \quad \quad \leq C_{N,n}  (1+t)^{-n-2} \|f_1\|_{E_{(1), N'}}^{2},
\end{split}
\end{equation}
ii) if $F$ is given by (\ref{a1.7.2}) then, %
$C_{N,n}$ only depends on $\|f \|_{E_{ 3}}$ and
\begin{equation} \label{Eq2.9}
\begin{split}
&q_N\left(\frac{d^n}{dt^n}
\left(-e^{-i \epsilon \omega_{m}(-i\nabla)t}
       (2i\omega_{m}(-i\nabla))^{-1}a^{(+)}_{1,-\epsilon}(t))
      (2i\omega_{2m}(-i\nabla))^{-1}a^{(+)}_{2,\epsilon}(t))
           - \dot{b}^{(+)}_{1,\epsilon} (t)\right)\right) \\
& \quad \quad \leq C_{N,n}  (1+tm)^{C \|f_2 \|_{E_{(2),1}} -n-2} (\ln{(2+tm)})^{2N'+1}\|f_1\|_{E_{(1), N'}} \|f_2\|_{E_{(2), N'}}.
\end{split}
\end{equation}
\end{lemma}
\textbf{Proof:} We only consider the case (ii). Define $h,$ $I$ and $J$ by
\begin{equation} \notag %
\begin{split}
&(h(s))^{\hat{}}(k)=\frac{i\epsilon}{4m}
(b^{(+)}_{1,-\epsilon}(s))^{\hat{}} (-k) f_{2,\epsilon}^{\hat{}}(2k), \;\;
I(t)=\dot{b}^{(+)}_{1,\epsilon} (t)-h(t)/t, \;\;
J(t,s)=- h (s)/t \\ &-e^{-i \epsilon \omega_{m}(-i\nabla)t}
       ((2i\omega_{m}(-i\nabla))^{-1}e^{-i \epsilon \omega_{m}(-i\nabla)t}b^{(+)}_{1,-\epsilon}(s))
      ((2i\omega_{2m}(-i\nabla))^{-1}e^{-i \epsilon \omega_{2m}(-i\nabla)t}f_{2,\epsilon}).
           \end{split}
\end{equation}
We have to prove that
$q_N (\frac{d^n}{dt^n} (J(t,t)-I(t)))$ is majorized by the right hand side of
inequality (\ref{Eq2.9}). Let $J_{n_1,n_2}(t,s)=(d/(dt))^{n_1}(d/(ds))^{n_2} J(t,s).$
Theorem~\ref{th 3.3} (with $n=0$ and $(d/(ds))^{n_2}h(s)$ instead of $f_0$),
gives
\begin{equation} \notag %
q_N (J_{n_1,n_2}(t,s)) \leq C_{N,n}
(1+tm)^{-n_1-2} q_{N'}((\frac{d}{ds})^{n_2}b^{(+)}_{1,\epsilon} (s)) \,
               \|f_2\|_{E_{(2), N'}}.
\end{equation}
Inequality (\ref{Eq2.7}) then gives, %
 with $n=n_1+n_2$ and  new $C_{N,n}$ and $N':$
\begin{equation} \notag %
q_N (J_{n_1,n_2}(t,t)) \leq  C_{N,n} (1+mt)^{C \|f_2 \|_{E_{(2),1}} -n-2}
(\ln{(2+tm)})^{2N'+1}      \|f_1\|_{E_{(2), N'}}  \|f_2 \|_{E_{(2), N'}}.
\end{equation}
Summing over $n_1+n_2=n,$ it follows that $q_N (\frac{d^n}{dt^n} J(t,t))$
is majorized by the right hand side of inequality (\ref{Eq2.9}).
This is also the case for  $q_N (\frac{d^n}{dt^n} I(t,t)).$ In fact, according to
(\ref{Eq6 proof lm1}),
$(I(t))^{\hat{}}(k)=(i \epsilon / (mt))((1+\omega_m(k)/(tm^2))^{-1}-1)
(b^{(+)}_{1,-\epsilon}(s))^{\hat{}} (-k) f_{2,\epsilon}^{\hat{}}(2k).$
Derivation in $t$ and application of inequality (\ref{Eq2.7}) now give the result.
\qed %

To state the main result on the existence of covariant modified wave operators for equation
(\ref{a1.6}), with  the nonlinearities (\ref{a1.7.1}) and (\ref{a1.7.2}),
we define $\mathcal{O}^{+} =E_\infty$ in the case of (\ref{a1.7.1}) and
$\mathcal{O}^{+} =\{f \in E_\infty \; | \; C \|f_2 \|_{E_{(2),1}} <1\}$
in the case of (\ref{a1.7.1}), where $C>0$ is as in Lemma \ref{lm2}.
\begin{theorem} \label{th2} If $f \in \mathcal{O}^{+} $ then, there
exists a unique solution $a \in C(\mathbb{R}, (I-\Delta)^{-1}E)$ of equation (\ref{Eq1.9}), such that
the asymptotic condition (\ref{Eq1.5.3}) is satisfied with $\alpha =0$.
This solution satisfies (\ref{Eq1.5.3}) for an $\alpha >0$ and
$a \in C^\infty (\mathbb{R}, E_\infty)$ and defines by (\ref{Eq1.5.4}) a $C^{\infty}$
modified wave operator $\Omega_+ : \mathcal{O}^{+} \rightarrow E_\infty.$
$\Omega_+$ intertwines the linear and nonlinear representations of $\mathcal{P},$
i.e. for all $f \in \mathcal{O}^{+}$ there exists a neighborhood of the identity in $\mathcal{P}$
of elements  $g$ such that $U_g (\Omega_+(f))=\Omega_+(U^1_gf).$
\end{theorem}
\textbf{Proof:} We only consider the case of the nonlinearity (\ref{a1.7.2}),
since the case (\ref{a1.7.1}) is easier. Let $f \in \mathcal{O}^{+}.$
For $j=1,2,$
$T^{2}_{(j)}$ and $T^{2}_{(j,\epsilon)}$ be the orthogonal projections of $T^{2}$
on $E_{(j)}$ and  $E_{(j,\epsilon)}$ respectively.
We shall use the following notations, where  $g, h(t) \in E_{(1),N},$ for some $N:$
\begin{equation} \label{Eq1 proof th2}
\begin{split}
&(H(h))(t)= -\int_t^\infty V_1(-s)T^{2}_{(1)P_0} (V(s)(h_1(s),f_2)) \, ds, \\
&I_{\epsilon}(t)= -\int_t^\infty \sum_{\epsilon_1+2\epsilon_2 \neq \epsilon}
V_{1,\epsilon}(-s)T^{2}_{(1,\epsilon)P_0} (V_{1,\epsilon_1}(s) (b^{(+)}_{1,\epsilon_1}(f))(s),
              V_{2,\epsilon_2}(s)f_{2,\epsilon_2}) \, ds, \\
&J_\epsilon (t)= -\int_t^\infty
\big(
V_{1,\epsilon}(-s)T^{2}_{(1,\epsilon)P_0} (V_{1,-\epsilon}(s)(b^{(+)}_{1,-\epsilon}(f))(s),
            V_{2,\epsilon}(s)f_{2,\epsilon}(s))) 
     -(\dot{b}_{1,\epsilon}^{(+)}(f))(s)
       \big) \, ds, \\
&(K_\epsilon (g,f_2))^{\hat{}}(k)= \frac{i}{2 \pi}\int_{\mathbb{R}^2} \sum_{\epsilon_1+2\epsilon_2 \neq \epsilon}
d_{\epsilon,\epsilon_1,\epsilon_2}(p,k-p) \frac{\hat{g}_{\epsilon_1}(p)}{2i\omega_{m}(p)}\frac{\hat{f}_{2,\epsilon_2}(k-p)}{2i\omega_{2m}(k-p)}, \\
&d_{\epsilon,\epsilon_1,\epsilon_2}(p_1,p_2) =(\epsilon \omega_{m}(p_1+p_2)-\epsilon_1 \omega_{m}(p_1)-\epsilon_2 \omega_{2m}(p_2))^{-1}.
\end{split}
\end{equation}
Given $c >0$ let $M_{\tau},$ where $\tau>0,$ be the Banach space of functions $h \in C([\tau, \infty[, (I-\Delta)^{-1}E_{(1)})$
with norm $|||h|||=\sup_{t \geq \tau} (1+t)^c \|(I-\Delta)h(t)\|_{E_{(1)}}  <\infty.$ Using that
$\|V_2(t)(\omega_{2m}(-i\nabla))^{-1/2}f_2\|_{L^{\infty}}
  \leq C' (1+t)^{-1} \|f_2\|_{E_{(2)N_0}}$ for some $N_0$ it follows that %
$|||H(h)||| \leq C_\tau \, |||h ||| \; \|f\|_{E_{(2)N'}}$ for some $N'$ and $C_\tau.$
To estimate $J,$ for the given $f \in \mathcal{O}^{+}$ we choose $c$ such that
$0 <c <1-C \|f_2 \|_{E_{(2),1}}.$ Inequality (\ref{Eq2.9}) of Lemma \ref{lm2},
with $N=2$ and $n=0,$ then gives  that $||| J ||| \leq C' \|f_1\|_{E_{(1), N'}} \|f_2\|_{E_{(2), N'}}$
for some new $N'$ and $C'.$ To estimate the non-resonant terms $I(t)$ we
proceed, with minor changes, as in \S3 of \cite{S-T 92}. We obtain (see Corollary 3.8 of \cite{S-T 92})
$\|(I-\Delta)K(V_1(t)g,V_2(t)f_2)\|_{E_{(1)}}
  \leq C' (1+t)^{-1} \|g\|_{E_{(1)N_0}} \|f_2\|_{E_{(2)N_0}}$ for some $C',$ $N_0.$
Partial integration gives
\begin{equation} \label{Eq2 proof th2}
I(t)=K(V_1(t)(b^{(+)}_{1}(f))(t),V_2(t)f_2)+\int_t^\infty K(V_1(s)(\dot{b}^{(+)}_{1}(f))(s),V_2(s)f_2) \, ds
\end{equation}
By Lemma \ref{lm1} we now obtain (with new constants) that $||| I ||| \leq C' \|f_1\|_{E_{(1), N'}} \|f_2\|_{E_{(2), N'}}.$
These estimates give, with $ G(h,f)=H(h)+I+J$ that
$|||G(h,f)||| \leq C' ( |||h||| \; \|f_2\|_{E_{(2), N'}}  +  \|f_1\|_{E_{(1), N'}} \|f_2\|_{E_{(2), N'}})$
and $|||G(h,f)-G(h',f)||| \leq C'  |||h-h'||| \; \|f_2\|_{E_{(2), N'}}.$ Let $f_2$ be such
that $C'  \; \|f_2\|_{E_{(2), N'}} <1,$ so $G$ is a contraction. The equation
(\ref{Eq1.9}) for $t \geq \tau$ is, since $(b_2(f))(t)=f_2,$ equivalent to
\begin{equation} \label{Eq3 proof th2}
b_1- b^{(+)}_1(f)=G(b_1- b^{(+)}_1(f),f). %
\end{equation}
This equation has a unique solution $b- b^{(+)}(f) \in M_\tau.$ It follows using Gr\"onwall's lemma that,
 there is a unique continuation of $V(\cdot)b(\cdot)$ to a solution
$a \in C(\mathbb{R},(I-\Delta)^{-1}E_{})$ of the integrated version of (\ref{a1.6.1})
and that $b_1- b^{(+)}_1(f) \in M_0.$
Similarly, one establish that the mappings
$\mathcal{O}^{+} \ni f \mapsto b(0)=a(0)=\Omega_+(f) \in (I-\Delta)^{-1}E_{}$
and $f \mapsto b,a \in M_\tau,$ $\tau \in \mathbb{R}$ are $C^\infty.$

We next turn to the covariance properties of $\Omega_+.$ For given $f \in \mathcal{O}^{+},$
we consider an open neighborhood of the identity of elements $g \in \mathcal{P}$ such that
$U^1_g f \in \mathcal{O}^+.$ $a^g$ denotes the solution in $C(\mathbb{R},(I-\Delta)^{-1}E_{}),$
given by the above construction, of (\ref{Eq1.9}) with scattering data $U^1_g f$
and  $b^g (t) =V(-t)a^g(t).$ Equation (\ref{Eq1.9}) gives
\begin{equation} \label{Eq4 proof th2}
b^g (t)- (b^{(+)}(U^1_g f))(t)=
 \int_\infty^t (V(-s) T^{2}_{P_0} (V(s)b^g(s))
           -(\dot{b}^{(+)}(U^1_gf))(s) ) \, ds.
\end{equation}
Let $R'$ be the representation of $\mathcal{P},$ on tempered distributions
$F \in S'(\mathbb{R}^3,\mathbb{C}^4),$
defined by the representation $R$ in (\ref{a1.7.3}) on $\phi_1,\phi_2 \in S'(\mathbb{R}^3,\mathbb{C})$
and the transformations (\ref{a1.4}) and $F(t)=V(-t)a(t).$
For translations, i.e. $g=(I,(s_0,s_1,s_2)),$ formula (\ref{Eq1 prooof lm1})
shows that $b^{(+)}(U^1_g f)=U^1_g  b^{(+)}(f).$
 Let $\delta_g(f)=R'_g b^{(+)}(f)-b^{(+)}(U^1_g f).$
If $g$ is a space translation, i.e. $s_0=0,$  then $(R'_g F)(t,x)=F(t,x_1+s_1,x_2+s_2),$
so $\delta_g(f)=0.$ %
If $g$ is a time translation, i.e. $s_1=s_2=0,$ then $(R'_g F)(t,x)=V(s_0)F(t+s_0,x),$
so $(R'_g b^{(+)}(f))(t)=(b^{(+)}(U^1_g f))(t+s_0)$ and
$(\delta_g(f))(t)=V(s_0)( (b^{(+)}(f))(t+s_0)-(b^{(+)}(f))(t)).$ With $c$ as above,
let $0 <c'<c.$ Then Lemma \ref{lm2} gives that
$\|\frac{d}{dt} (\delta_g(f))(t) \|_{E_N} \leq |s_0| C'_N (1+t)^{-c'-1},$
for some $C'_N.$ We can now integrate $\frac{d}{dt} (\delta_g(f))(t)$ in formula
(\ref{Eq4 proof th2}), which shows that for sufficiently small translations
\begin{equation} \label{Eq5 proof th2}
b^g (t)- (R'_g b^{(+)}(f))(t)=
 \int_\infty^t (V(-s) T^{2}_{P_0} (V(s)b^g(s))
           -\frac{d}{ds}(R'_g b^{(+)}(f))(s) ) \, ds.
\end{equation}
It follows that (\ref{Eq5 proof th2}) holds true with $b^g=R'_g b$
and this solution is unique. If $g$ is a space rotation, then similarly one finds
that $b^g=R'_g b.$ This shows the intertwining property, $U_g(\Omega_+(f))=\Omega_+(U^1_gf),$ for
$g$ in a neighborhood of the identity in the $\mathbb{R}^3 \psd SO(2)$ subgroup of $\mathcal{P}.$

For the case of a Lorentz transformation, let $g(s)=\exp{(s N_{j})},$ $j=1,2$
and $X(t)=N_{j}+tP_j.$ The already proved intertwining property shows that
$a^{g(s)}(t)=\Omega_+(V(t)U^1_{g(s)}f).$ This function is $C^\infty$ in $(t,s,f),$
since $\Omega_+$ is  $C^\infty.$ Suppose for the moment that, for $t=s=0,$
\begin{equation} \label{Eq6 proof th2}
\frac{d}{ds}a^{g(s)}(t)=T_{X(t)}(a^{g(s)}(t)).
\end{equation}
Then one obtains
$D\Omega_+(f;T^1_{N_{j}}f)= T_{N_{j}}(a^{g(0)}(0))=T_{N_{j}}(\Omega_+ (f)),$
which shows that the  intertwining property holds true for a neighborhood of the identity
in $\mathcal{P}.$ Successive differentiation of $g \mapsto \Omega_+(U^1_{g(s)}f) \in (I-\Delta)^{-1}E_{}$
gives that $T_Y(\Omega_+(f)) \in E_{}$ for all $Y \in \Pi'.$ Now according to Theorem 2 of \cite{S-T 95},
$\Omega_+(f) \in E_{\infty}$ and this mapping from $\Omega_+$ to $E_\infty$ is $C^\infty.$

To complete the proof we shall prove formula (\ref{Eq6 proof th2}) for $t\geq 0.$
The differentiability  of $b$ in $f,$ justifies to differentiate in $s$ both sides
of formula (\ref{Eq4 proof th2}), with $g=g(s).$ %
Then, with $b^{g(0)}=b$ and $b'=db^{g(s)}/(ds)|_{s=0}$ and
$b^{'(+)}=d b^{(+)}(U^1_g f)/(ds)|_{s=0}:$
\begin{equation} \label{Eq7 proof th2}
b' (t)- b^{'(+)}(t)= \int_\infty^t
  (V(-s) 2(DT^{2}_{P_0})(V(s)b(s);V(s)b'(s))
           -\frac{d}{ds}b^{'(+)}(s) ) \, ds.
\end{equation}
The generator $\Xi_{N_{j}}$ of $s \mapsto R'_{g(s)},$ is given by its component in $E_{(1,\epsilon)}:$
\begin{equation} \label{Eq8 proof th2}
((\Xi_{(1,\epsilon)N_{j}}h)(t))^{\hat{}}(k)=
\big(-\epsilon (\omega_m(k)\frac{\partial}{\partial k_j} + \frac{k_j t}{\omega_m(k)} \frac{\partial}{\partial t})
  +i(\frac{\partial}{\partial k_j}-\frac{k_j }{(\omega_m(k))^2})\frac{\partial}{\partial t}\big)
               (h(t))^{\hat{}}(k).
\end{equation}
Using Lemmas \ref{lm1} and \ref{lm2} one establish, with $c'$ as above, that
$\|\frac{d}{dt} \big( (\Xi_{N_{j}}b^{(+)})(t) -b^{'(+)}(t)\big)\|_{E_N} \leq |s_0| C'_N (1+t)^{-c'-1},$
for some $C'_N.$
This shows that we can replace $b^{'(+)}$ by $\Xi_{N_{j}}b^{(+)}$ on both sides of (\ref{Eq7 proof th2}).
Observing that $V(t)(\Xi_{N_{j}}b)(t)=T_{X(t)}(a(t))$
and that $T^{1}_{P_0}V(t)(\Xi_{N_{j}}b)(t)+2(DT^{2}_{P_0})(V(t)b(t);V(t)(\Xi_{N_{j}}b)(t))=T_{P_0 X(t)}(a(t))$
we can now identify $b'(t)$ with $V(-t)T_{X(t)}(a(t))$ which satisfies the equality
\begin{equation} \notag %
V(-t)T_{X(t)}(a(t))- (\Xi_{N_{j}}b^{(+)})(t)= \int_\infty^t
  (V(-s) 2(DT^{2}_{P_0})(a(s);T_{X(t)}(a(s)))
           -\frac{d}{ds} (\Xi_{N_{j}}b^{(+)})(s) ) \, ds.
\end{equation}
\qed
\section{The linear K-G equation} \label{lin k-g}
We shall here give certain results on phase and decrease properties of solutions
of linear Klein-Gordon equations, which we have used to study resonant terms.
They are adapted from  Appendix \cite{FST97} to our situation and are based on the
symbolic calculus developed in \cite{Horm87}. Let $M >0.$ For given $\epsilon =\pm$ and
$f \in S(\mathbb{R}^{2},\mathbb{C}),$
\begin{equation}
 \label{Eq3.1}
 \varphi (t)= e^{i \epsilon \omega_{M}(-i\nabla)t}f,
\end{equation}
defines a  solution $\varphi$ of
\begin{equation}
 \label{Eq3.2}
(\square + M^{2}) \varphi=0.
\end{equation}
The forward light-cone is denoted
$\Gamma_+=\{(t,x) \in \mathbb{R}^3 \, \big\vert \, t^2-|x|^2 \geq 0, t\geq 0 \}$
and let $\rho = (t^2-|x|^2)^{1/2}$ for $(t,x) \in \Gamma_+.$
The sequence of functions
$g^{}_l \in C^\infty ((\mathbb{R}^+ \times \mathbb{R}^2)- \{ 0 \}),$
with support in the forward light-cone,
is defined by %
\begin{equation} \label{Eq3.3}
g^{}_0 (t,x) = i \epsilon  (Mt/\rho^2)  \hat{f}(-\epsilon M x / \rho),  \; \;
g^{}_l = \frac{\rho}{ 2i \epsilon l M}  \square g^{}_{l-1},
\quad l \geq 1,
\end{equation}
for $(t,x) \in \Gamma_+.$
$g^{}_l$ is homogeneous of degree $- 1 - l$. The solution $\varphi=\varphi_0$
has an asymptotic expansion with rest-term $\varphi_n:$
\begin{equation} \label{Eq3.4}
\varphi_n =\varphi_0-e^{i\epsilon M \rho}\sum_{0\leq l\leq n-1}g^{}_l,
\quad n\geq 1.
\end{equation}
Define $\lambda(t)$ and $\delta(t)$ for $t\geq 0$ by
\begin{align} \label{Eq3.5}
(\lambda(t))(x)&= t/(1+t-\vert x\vert)\quad \hbox{for}\ 0\leq
\vert x\vert \leq t, \\
(\lambda(t))(x)&=\vert x\vert \quad \hbox{for}\ 0\leq t \leq
\vert x\vert, \\
(\delta(t))(x)&=1+t+\vert x\vert.
\end{align}
We introduce the representation $X\mapsto \xi^{}_X$ of the Poincar\'e Lie
algebra $\mathfrak{p}$ by:
\begin{align} \label{Eq3.6}
\xi^{}_{P_0}&=\frac{\partial}{\partial t}, \quad
\xi^{}_{P_i}=\frac{\partial}{\partial x^{}_i},\quad 1\leq i\leq 2, \\
\xi^{}_{N_{i}}&=x^{}_i \frac{\partial}{\partial t}
+t \frac{\partial}{\partial x^{}_i},\quad 1\leq i\leq 2, \\
\xi^{}_{R}&=-x^{}_1 \frac{\partial}{\partial x^{}_2}
+x^{}_2 \frac{\partial}{\partial x^{}_1}.
\end{align}
We define, for a function $d: \Pi' \rightarrow S(\mathbb{R}^{2},\mathbb{C})$
and for $n \in \mathbb{N}:$
\begin{equation} \label{Eq3.7}
p^{(s)}_n (d) = \sum_{\substack{Y\in \Pi' \\ \vert Y\vert \leq n}}
(M\Vert  d_Y \Vert_{L^s} +\sum_{0 \leq \mu \leq 2}\Vert  d_{P_{\mu}Y} \Vert_{L^s}).
\end{equation}

The following theorem  gives decrease properties
of the solution $\varphi$ and the rest terms $\varphi_n.$ We omit its proof,
since its so similar to that of Theorem A.1 in \cite{FST97}, considering the case of
the Dirac equation in $3$-space dimensions. Given an ordering on the basis
$Q=\{N_{1}, N_{2},R\}$ of ${\mathfrak{so}}(2,1),$ let $Q'$ be the corresponding
standard basis of the enveloping algebra $U({\mathfrak {so}}(2,1))$ of ${\mathfrak {so}}(2,1)$.
\begin{theorem} \label{th 3.1}
There exists  $C_i \in \mathbb{R},$ $i\geq0,$ and $\kappa_i \in \mathbb{N},$ $1 \leq i\leq4,$
such that for all
$j,k,n\in\mathbb{N}$, $t >0$, $f \in S(\mathbb{R}^{2},\mathbb{C}),$
$X\in\Pi'\cap U(\mathbb{R}^3)$ and $Y\in Q':$
\begin{align} \label{Eq3.8}
&p^{(2)}_j\big((1+\lambda(t))^{k/2}(\xi  \varphi_0)(t)\big)
\leq C_{j+k}\big( \bar{q}_{j+k}(f)+\sum_{1\leq i\leq 2} \bar{q}_{j+k}(\partial_i f)\big), \\
 \label{Eq3.9}
&p^{(\infty)}_j\big(\delta(t)(1+\lambda(t))^{k/2}(\xi \varphi_0)(t)\big)
\leq C_{j+k} \bar{q}_{j+k+\kappa_2}(f)
, \\
 \label{Eq3.10}
&p^{(2)}_j\big((1+\lambda(t))^{k/2}(\xi  \varphi_{n+1})(t)\big)
\leq C_{j+k+n}\ t^{-n-1} \bar{q}_{3(j+k+n)+\kappa_1} (f), \\
 \label{Eq3.11}
&p^{(\infty)}_j\big(\delta(t)(1+\lambda(t))^{k/2}(\xi \varphi_{n+1})(t)\big)
\leq C_{j+k+n}\ t^{-n-1}\bar{q}_{3(j+k+n)+\kappa_3} (f), \\
 \label{Eq3.12}
&\Vert (\rho^{-j}\xi^{}_{XY}g^{}_n)(t)\Vert^{}_{L^2}
\leq C_{j+\vert X\vert +n}\ t^{-j-n-\vert X\vert}
\bar{q}_{3\vert X\vert+3n+\vert Y\vert+j} (f), \\
 \label{Eq3.13}
&\Vert (\rho^{-j}\xi^{}_{XY}g^{}_n)(t)\Vert^{}_{L^\infty}
\leq C_{j+\vert X\vert+ n}\ t^{-1-j-n-\vert X\vert}
\bar{q}_{3\vert X\vert+3n+\vert Y\vert+j+\kappa_4} (f).
\end{align}
\end{theorem}
The development defined by (\ref{Eq3.3}) and (\ref{Eq3.4}) can be inverted.
Given a homogeneous function
$g \in C^\infty ((\mathbb{R}^+ \times \mathbb{R}^2) - \{0\})$ of degree $- 1$
with support in $\Gamma_+,$
we construct by iteration $f_0,\ldots,f_n \in D_\infty:$
\begin{align} \label{Eq3.14}
&\hat{f}_l (k)= -i \epsilon (M/ (\omega(k))^2)  g^{}_{l,0} (1,- \epsilon k /\omega_M(k)),
\quad 0 \leq l \leq n, \\
\label{Eq3.15}
& g^{}_{0,0} = g,\quad  g^{}_{l,0} (t,x) = - \sum_{1 \leq j \leq l}  t^j
g^{}_{l-j,j} (t,x),\quad  1 \leq l \leq n, \\
\label{Eq3.16}
&g^{}_{l,j}= \frac{\rho}{ 2i \epsilon j M}  \square g^{}_{l,j-1},\quad  1 \leq j
\leq n-l.
\end{align}
By this construction $g^{}_{l,j} \in C^\infty ((\mathbb{R}^+ \times \mathbb{R}^2) - \{0\})$ is
homogeneous of degree $- 1-j$ with support in $\Gamma_+.$ Reformulation
in two space dimensions of Theorem A.2 \cite{FST97},
(there proved in the case of three space dimensions), gives:
\begin{theorem} \label{th 3.2}
Let $g \in C^\infty ((\mathbb{R}^+ \times \mathbb{R}^2) - \{0\})$ be a
homogeneous function of degree $-1$ with ${\rm supp}\ g \subset \Gamma_+.$
If $f_0,\ldots,f_n$ are given by the construction (\ref{Eq3.14})-(\ref{Eq3.16}) and
\begin{equation}
 \label{Eq3.16.1}
u^{}_n(t)=e^{i \epsilon M \rho }g(t)
-\sum_{0\leq l\leq n} t^{-l}e^{i\epsilon\omega_M(-i\nabla)t}f^{}_l,
\end{equation}
then there exists  $C_i \in \mathbb{R},$ $i\geq0,$ independent of $g,$
such that for all
$j,k,n\in\mathbb{N}$ and  $t >0$ and with $\kappa_1$ and $\kappa_3$ as in Theorem \ref{th 3.1}:
\begin{align}
\label{Eq3.17}
& \bar{q}_j(f^{}_n) \leq C_{n+j}  \sum_{\substack{ Y\in Q' \\ q+\vert Y \vert \leq j+2n}}
\Vert ({m / \rho(1,\cdot)})^{q+n}  (\xi^{}_Y g) (1,\cdot) \Vert^{}_{L^2}, \\
\label{Eq3.18}
&p^{(2)}_j\big((1+\lambda(t))^{k/2}(\xi u_n) (t)\big)
\leq C_{j+k+n}\sum_{\substack{ Y\in Q' \\ q+\vert Y \vert \leq 3(j+k+n)+\kappa_1}}
\Vert ({m / \rho(1,\cdot)})^q  (\xi^{}_Y g) (1,\cdot) \Vert^{}_{L^2}t^{-n-1}, \\
\label{Eq3.19}
&p^{(\infty)}_j\big(\delta(t)(1+\lambda(t))^{k/2} (\xi u_n) (t)\big)
\leq C_{j+k+n}\sum_{\substack{ Y\in Q' \\ q+\vert Y \vert \leq 3(j+k+n)+\kappa_3}}
\Vert ({m / \rho(1,\cdot)})^q  (\xi^{}_Y g) (1,\cdot) \Vert^{}_{L^2}t^{-n-1}.
\end{align}
\end{theorem}
Theorem \ref{th 3.1} and Theorem \ref{th 3.2} permit to find the asymptotic behavior
and estimates of resonant terms.
\begin{theorem} \label{th 3.3}
Let $M, M_1, M_2 >0$ and $\epsilon, \epsilon_1, \epsilon_2 \in \{- 1,1\}$
be such that $\epsilon M=\epsilon_1 M_1 + \epsilon_2 M_2$ and let
$f^{(1)},f^{(2)} \in S(\mathbb{R}^{2},\mathbb{C}).$
There exists a unique
sequence of functions $f_{l} \in S(\mathbb{R}^{2},\mathbb{C}),$ such that if
\begin{equation}
 \label{Eq3.20}
\delta_n (t)=e^{-i\epsilon\omega_M(-i\nabla)t}
\left((e^{i\epsilon_1\omega_{M_1}(-i\nabla)t}f^{(1)}) (e^{i\epsilon_2\omega_{M_2}(-i\nabla)t}f^{(2)})\right)
-\sum_{0\leq l\leq n} t^{-1-l}f^{}_l,
\end{equation}
then for all $N,j,n \in \mathbb{N}$ there are $C$ and $N'$ such that
\begin{equation}
 \label{Eq3.21}
 \bar{q}_N((\frac{d}{dt})^j \delta_n (t)) \leq C (1+t)^{-2-j-n}  \bar{q}_{N'}(f^{(1)}) \bar{q}_{N'}(f^{(2)}).
\end{equation}
Moreover
\begin{equation}
 \label{Eq3.22}
\hat{f}_0(k)= i \frac{\epsilon_1 M_1 \epsilon_2 M_2}{\epsilon M} (\frac{\omega_{M}(k)}{M})^2
(f^{(1)})^{\hat{}}(\frac{\epsilon_1 M_1}{\epsilon M}k)
(f^{(2)})^{\hat{}}(\frac{\epsilon_2 M_2}{\epsilon M}k)
\end{equation}
and
\begin{equation}
 \label{Eq3.23}
\bar{q}_N (f_j) \leq  C  \bar{q}_{N'}(f^{(1)}) \bar{q}_{N'}(f^{(2)}).
\end{equation}
\end{theorem}
\textbf{Proof:}
With $f^{(i)}$ instead of $f,$ we define $\varphi^{(i)},$ $g_n^{(i)}$ and $\varphi_n^{(i)}$
by formulas (\ref{Eq3.1})--(\ref{Eq3.4}). Given $J \in \mathbb{N},$ formula (\ref{Eq3.20})
can be written
\begin{equation}
 \label{Eq3.24}
\begin{split}
\delta_n (t)=&e^{-i\epsilon\omega_M(-i\nabla)t}
(
(\varphi_{J+1}^{(1)}(t)+e^{i\epsilon_1 M_1 \rho}\sum_{0 \leq l \leq J}g_l^{(1)}(t))
(\varphi_{J+1}^{(2)}(t)+e^{i\epsilon_2 M_2 \rho}\sum_{0 \leq l \leq J}g_l^{(2)}(t))
) \\
&-\sum_{0\leq l\leq n} t^{-1-l}f^{}_l,
\end{split}
\end{equation}
where the functions $f^{}_l$ will be defined later in this proof. We define
\begin{equation}
 \label{Eq3.25}
g_l(t, \cdot)=t^{1+l} \sum_{\substack{l_1+l_2=l \\ 0 \leq l_i \leq J}}
      g^{(1)}_{l_1}(t, \cdot) g^{(2)}_{l_2}(t, \cdot),
\end{equation}
\begin{equation}
 \label{Eq3.26}
I_J(t)=\varphi_{J+1}^{(1)}(t) \varphi_{J+1}^{(2)}(t)
   +\sum_{0 \leq l \leq J}
   (e^{i\epsilon_1 M_1 \rho}g_l^{(1)}(t) \varphi_{J+1}^{(2)}(t)
     +e^{i\epsilon_2 M_2 \rho}g_l^{(2)}(t) \varphi_{J+1}^{(1)}(t))
\end{equation}
and
\begin{equation}
 \label{Eq3.27}
v_n^J(t)=
e^{-i\epsilon\omega_M(-i\nabla)t}\sum_{0\leq l\leq 2J} t^{-1-l} e^{i\epsilon M \rho}g_l(t, \cdot)
-\sum_{0\leq l\leq n} t^{-1-l}f^{}_l.
\end{equation}
Then
\begin{equation}
 \label{Eq3.27.1}
\delta_n(t)= 
e^{-i\epsilon\omega_M(-i\nabla)t}I_J(t) +v_n^J(t).
\end{equation}
The function $g_l$ is homogeneous of degree $-1.$
We note that, according to (\ref{Eq3.12}) and (\ref{Eq3.13}) of Theorem \ref{th 3.1},
if $Z \in \Pi'$ then
\begin{equation}
 \label{Eq3.28}
\Vert (\rho^{-j}\xi^{}_{Z}g^{}_l)(1, \cdot)\Vert^{}_{L^2}
\leq C \bar{q}_{N'} (f^{(1)})\bar{q}_{N'} (f^{(2)}),
\end{equation}
where $C$ and $N'$ depend on $|Z|$ and $j.$ Also, a straight forward application
of (\ref{Eq3.10})-- (\ref{Eq3.13}) gives, with new $C$ and $N'$ depending on
$N,$ $k$ and $J$
\begin{equation}
 \label{Eq3.29}
p^{(2)}_N \big((1+\lambda(t))^{k/2}(\xi I_J)(t)\big)
\leq  C  t^{-2-J}  \bar{q}_{N'}(f^{(1)})\bar{q}_{N'} (f^{(2)}).
\end{equation}

With $g_l$ instead of $g$ and $J$ instead of $n,$ we define $f_{l,k}$ and $u_{l,J}$
by formulas (\ref{Eq3.14})--(\ref{Eq3.16.1}).
Then, according to Theorem \ref{th 3.1} and (\ref{Eq3.28}),
\begin{equation}
 \label{Eq3.30}
u_{l,J}(t)=e^{i \epsilon M \rho }g_l(t)
-\sum_{0\leq k\leq J} t^{-k}e^{i\epsilon\omega_M(-i\nabla)t}f_{l,k}
\end{equation}
satisfies, with $C$ and $N'$ depending on $J,$ $N$ and $k,$
\begin{equation}
\label{Eq3.31}
\begin{split}
t^{J+1}&
  \left(   p^{(2)}_N\big((1+\lambda(t))^{k/2}(\xi u_{l,J}) (t)\big)
    + p^{(\infty)}_N\big(\delta(t)(1+\lambda(t))^{k/2}(\xi u_{l,J}) (t)\big) \right) \\
  &+\bar{q}_N (f_{l,k})
                \leq  C \bar{q}_{N'}(f^{(1)})\bar{q}_{N'} (f^{(2)}).
\end{split}
\end{equation}
Formulas (\ref{Eq3.27}) and (\ref{Eq3.30}) give
\begin{equation}
 \label{Eq3.32}
v_n^J(t)=\sum_{0\leq l\leq 2J} t^{-1-l}
\left(
e^{-i\epsilon\omega_M(-i\nabla)t} u_{l,J}(t) + \sum_{0\leq k\leq J} t^{-k} f_{l,k}
\right)
-\sum_{0\leq l\leq n} t^{-1-l}f^{}_l.
\end{equation}
In the sequel of this proof we suppose that $J \geq n$ and define
\begin{equation}
 \label{Eq3.33}
f_l=\sum_{k_1+k_2=l} f_{k_1,k_2}, \; A_J(t)=\sum_{0\leq l\leq 2J} t^{-1-l}u_{l,J}(t), \;
B_{n,J}(t)=\sum_{(l, k_1, k_2) \in D(n,J)} t^{-1-l}f_{k_1,k_2},
\end{equation}
where $ D(n,J)=\{(l, k_1, k_2) \, |  \, l>n, 0\leq l\leq 2J, 0\leq k_2 \leq J, k_1+k_2=l  \}.$
Then $v_n^J(t)=e^{-i\epsilon\omega_M(-i\nabla)t} A_J(t) + B_{n,J}(t),$ so by
(\ref{Eq3.27.1})
\begin{equation}
 \label{Eq3.34}
\delta_n(t)=
e^{-i\epsilon\omega_M(-i\nabla)t}h_J(t) + B_{n,J}(t), \; \;
\text{where} \:\; h_J(t)=I_J(t)+  A_J(t).
\end{equation}
Inequalities  (\ref{Eq3.28}) and  (\ref{Eq3.31}) and the fact that
$\xi_{N_i}t^{-1-l}=-(1+l)t^{-2-l}x_i$ give
\begin{equation}
\label{Eq3.35}
t^{J+2}   p^{(2)}_N\big((1+\lambda(t))^{k/2}(\xi A_J) (t)\big)
  +t^{n+j+2}\bar{q}_N ((\frac{d}{dt})^j B_{n,J}(t))
                \leq  C \bar{q}_{N'}(f^{(1)})\bar{q}_{N'} (f^{(2)}).
\end{equation}
Since
$(x_j i \epsilon  \omega_M(-i \nabla)-t\partial_j)  e^{-i \epsilon \omega_M(-i \nabla)t} =
e^{-i \epsilon \omega_M(-i \partial)t} x_j  i \epsilon  \omega_M(-i \nabla),$
it follows from Leibniz's rule and (\ref{Eq1.5.1.0}) that
\begin{equation}
\label{Eq3.36}
\begin{split}
&\bar{q}_N ((\frac{d}{dt})^j (e^{-i \epsilon \omega_M(-i \nabla)t} h_{J}(t)))
\leq C_{j,N}\sum_{\substack{j_1+j_2=j \\ |\alpha|, |\beta| \leq N}}
   \|x^{\alpha} \nabla^{\beta} e^{-i \epsilon \omega_M(-i \nabla)t}
      (\omega_M(-i \nabla))^{j_1} (\frac{d}{dt})^{j_2}  h_{J}(t) \|_{L^2}  \\
& \leq C'_{j,N}\sum_{\substack{|\alpha|+k \leq N \\ |\beta|+j_2 \leq N +j}}
  t^k \|x^{\alpha} \nabla^{\beta}  (\frac{d}{dt})^{j_2}  h_{J}(t) \|_{L^2}.
\end{split}
\end{equation}
Let $U(\mathbb{R}^3)$ be the enveloping algebra of the translation subalgebra of
$\mathfrak{p}.$  Inequalities (\ref{Eq3.29}), (\ref{Eq3.31}) and (\ref{Eq3.36})
give that for $t \geq 1:$
\begin{equation}
\label{Eq3.37}
\begin{split}
&\bar{q}_N ((\frac{d}{dt})^j (e^{-i \epsilon \omega_M(-i \nabla)t} h_{J}(t)))
\leq C''_{j,N}\sum_{\substack{X \in U(\mathbb{R}^3) \cap \Pi' \\ |X| \leq N+j}}
   t^N \|(1+\lambda (t))^N (\xi h_{J})(t) \|_{L^2} \\
& \leq C'''_{j,N} t^N p_{N+j}((1+\lambda (t))^N (\xi h_{J})(t))
 \leq C'_{j+N} t^{N-2-J}  \bar{q}_{N'}(f^{(1)})\bar{q}_{N'} (f^{(2)}).
\end{split}
\end{equation}
Inequality (\ref{Eq3.21}), for $t \geq 1,$ now follows by choosing $J \geq N+n+j.$
The definition of $f_l$ in formula (\ref{Eq3.33}) and inequality  (\ref{Eq3.31})
give inequality  (\ref{Eq3.23}).

To prove formula (\ref{Eq3.22}), we observe that
$g^{(j)}_0 (t,x) = i \epsilon_j  (M_j t/\rho^2)  \hat{f}(-\epsilon_j M_j x / \rho),$
$j=1,2,$ according to (\ref{Eq3.3}).
By (\ref{Eq3.25}), $g_0(t,x)=tg^{(1)}_0 (t,x)g^{(2)}_0 (t,x).$
By (\ref{Eq3.33}) $f_0=f_{0,0},$ so by formulas (\ref{Eq3.30})  and (\ref{Eq3.14})
$\hat{f}_0 (k)= -i \epsilon (M/ (\omega(k))^2)  g^{}_{0} (1,- \epsilon k /\omega_M(k)).$
The result now follows using that $\rho (1,- \epsilon k /\omega_M(k))=M/\omega_M(k).$
\qed \\

\noindent \textbf{Note added in the proofs:} I learned later from H.Sunagawa,
after the acceptation of the paper, that he has related results in
Hokkaido Math. Journ. \textbf{33}, 457--472 (2004), which cover some
of those for the easier case (\ref{Eq1.3}) but not for the case (\ref{Eq1.4}),
and the method is different.

\end{document}